\documentclass[aps,prl,
twocolumn,
showpacs]{revtex4}

\usepackage{amsmath}
\usepackage{amssymb}
\usepackage{graphicx} 
\usepackage{dcolumn}
\usepackage{bm}

\newcommand{\om}{\omega}

\newcommand{\br}{{\bf r}}

\newcommand{\tn}{{\widetilde{n} }}

\begin{document}

\title{

Stabilization of Solitons Generated by a Supersonic Flow
of Bose-Einstein Condensate Past an Obstacle}

\author{A. M. Kamchatnov$^{1}$}
\author{L. P. Pitaevskii$^{2,3,4}$}

\affiliation{$^1$Institute of Spectroscopy, Russian Academy of Sciences, Troitsk, Moscow Region, 142190, Russia\\
$^2$Dipartamento di Fisica, Universit\`a di Trento and CNR-INFM BEC Center, I-38050 Povo, Trento, Italy\\
$^3$Kapitza Institute for Physical Problems, ul. Kosygina 2, Moscow, 119334, Russia \\
$^4$Laboratoire Physique Th\'eorique et Mod\`eles Statistique, Universit\'e Paris Sud, B\^at. 100, 91405 Orsay CEDEX, France\\}

\date{\today}

\begin{abstract}
The stability of dark solitons generated
by supersonic flow of a Bose-Einstein condensate past an obstacle is investigated.
It is shown that in the reference frame
attached to the obstacle a transition occurs
at some critical value of the flow velocity from absolute
instability of dark solitons to their convective instability. This leads to the decay
of disturbances of solitons at a fixed
distance from the obstacle and the formation of effectively stable dark solitons.
This phenomenon explains the surprising stability of the flow picture that has been
observed in numerical simulations.
\end{abstract}


\maketitle

{\it 1. Introduction.}
It is well known that plane dark solitons are unstable with respect to
transverse (``snake'') perturbations. Such an instability was predicted and studied in
\cite{kp-1970,zakharov-1975} for the case of
shallow solitons described by the Kadomtsev-Petviashvili (KP) equation.
The instability of deep dark solitons in a Bose-Einstein
condensate (BEC), described by the Gross-Pitaevskii (GP) equation
was demonstrated in \cite{kt-1988}. These analytical predictions were
confirmed by both numerical simulations and experiments
with dark solitons in a
BEC (see, e.g., \cite{bca-2007} and references therein) and with optical solitons
(see,  \cite{ka-2003}).

It is worth noting that such soliton instability is not a property of particular equations,
but rather a general phenomenon of non-dissipative dynamics. Indeed,
as follows from \cite{KP04}, the motion of a 1D soliton can be described by the semi-classical
Newton equation $m_{s}d^2X/dt^2=F_{s}$, where $m_{s}=2dE_{s}/d(V^2)$ is the ``effective mass'' of the soliton,
$E_{s}$ is its energy, $V$ its velocity, and $F_{s}$ is the force acting on the soliton.
The crucial point is that in all relevant cases $E_{s}$ decreases with increasing velocity
and hence $m_{s}<0$. Let us apply now this equation to an element of a soliton deformed along the $y$ axis. If
the wavelength of the perturbation is long enough, the
soliton can be considered as a surface with surface tension $E_{s}$.
Then, according to the Laplace formula, the restoring surface force is equal to $F_{s}=E_{s}/R$, where
$R$ is the radius of curvature (see \cite{LL6}, \S61).  Let the equation of the surface
be $X=A\cos(py)$. In the linear approximation $R^{-1} \approx d^2X/dy^2=-p^2X$.
This results in an unstable dispersion relation for small oscillations:
\begin{equation}
   \omega=\pm i\left (E_{s}/|m_s| \right )^{1/2}p\;.
   \label{gen}
\end{equation}
In the GP equation case for which $E_{s} \propto (c^2-V^2  )^{3/2}$,
where $c$ is the speed of sound,
Eq.~(\ref{gen}) reduces to the equation $ \omega=\pm i\left [(c^2-V^2  )/3
\right ]^{1/2}p$ obtained in \cite{kt-1988}.

Dark solitons
in a uniform BEC are described by the GP equation  which we write here
in standard dimensionless units with $c=1$:
\begin{equation}\label{1-1}
   i \psi_t=-\tfrac{1}{2}\Delta\psi+
   |\psi|^2\psi,
\end{equation}
where $\psi(\br,t)$ is the condensate wave function. The solution  corresponding to
a plane dark soliton moving in the $x$ direction
was found in \cite{tsuzuki-1971} in the form ($k=\sqrt{1-V^2}$)
\begin{equation}\label{1-2}
    \psi(x+Vt)=
    [k     \tanh\left(k(x+Vt)\right)
    -iV]\exp(-i t)\;.
\end{equation}
The depth of a dark soliton depends on its
velocity $V$ which cannot exceed the sound velocity.
The dark soliton is unstable in the whole range of possible parameters.
Thus, dark solitons formed by means of density or phase engineering
\cite{bca-2007,bk-2003} are unstable
with respect to perturbations depending on the transverse
$y$ and $z$ coordinates.

However, there exists another possibility to generate dark solitons in a BEC.
As was shown in
\cite{egk-2006}, dark solitons can also be generated by a fast enough flow of
a BEC past an
obstacle as is illustrated in Fig.~1 where results of a
numerical simulation of a two-dimensional flow past a disk-shaped obstacle are shown.
\begin{figure}[t]
\includegraphics[width=7cm,height=6cm]{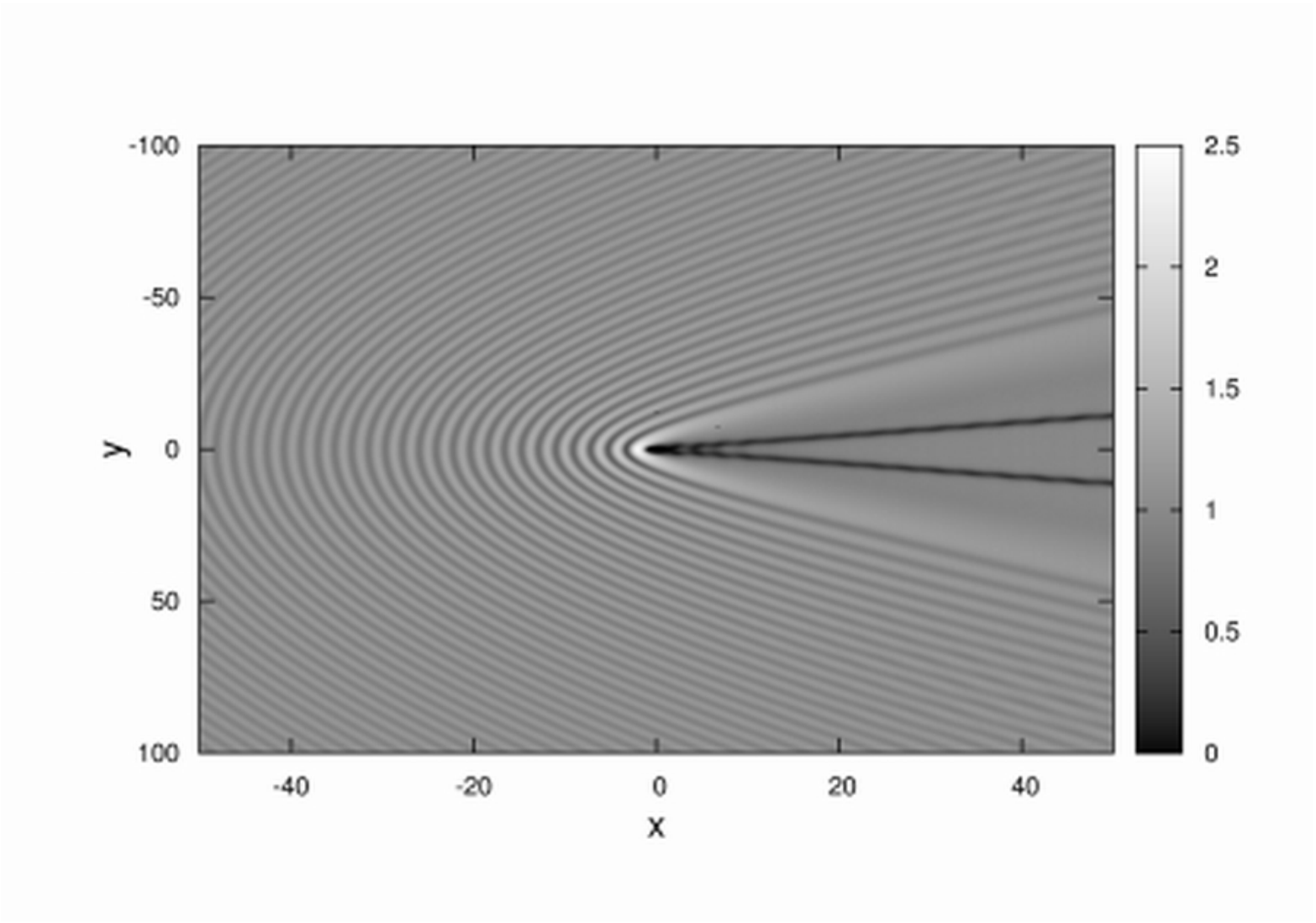}
\caption{Wave pattern generated by the flow of a BEC past an obstacle
located at the origin of the coordinate system.
The flow velocity is directed from left to right and corresponds to the
Mach number $M=2$.}
\label{fig1}
\end{figure}
We see that the Mach cone (imaginary lines drawn from the origin at angles
$\pm \arcsin(1/M)$ with respect to the $x$ axis, $M=2$ being the supersonic flow
velocity at $|x|\to\infty$) separates regions with wave patterns of
different nature. Outside the Mach cone there is a stationary
pattern arising due to waves with a Bogoliubov dispersion law radiated
by the obstacle. It is possible that these waves have been observed in recent
experiments \cite{cornell-2005,carusotto-2006}. Their theory has been developed in
\cite{carusotto-2006,gegk-2007,gk-2007}. Inside the Mach cone the nonlinear waves
are located. In case of large enough obstacles they form dispersive shock
waves considered in \cite{ek-2006}, but for obstacles with the size about a
healing length (about unity in our non-dimensional units), as in Fig.~1,
just one soliton is formed in each symmetrical ``shock''.
At large enough time of evolution and far enough from the obstacle,
the profile of a dark
soliton  tends to a stationary state described by the
solution (\ref{1-2}) of the GP equation \cite{egk-2006}. This means that
dark solitons generated by a fast enough flow of a BEC past an obstacle are stable
in the reference frame connected with the obstacle contrary to the  above mentioned
instability of dark solitons in the reference frame where the
fluid at infinity is at rest. We will see below that this seeming contradiction is
explained by the fact, that the instability of solitons in the
obstacle reference frame is ``convective'' only.
This means that disturbances are convected from
the region at finite distance behind the obstacle and, hence, the solitons
there become stable.

It is also known that at small enough velocity of the flow the situation is different.
Numerical solutions of the
GP equation demonstrate that a BEC flow past a cylindrical obstacle starts to generate
vortices at subsonic velocity $M\cong 0.45$ (see \cite{fpr-1992,wmca-1999,br-2000}).
However, the frequency of vortex
generation increases with increasing flow velocity. Therefore one can expect that
at large enough velocity vortices
are generated so often that the mean distance between them becomes less than their radial size
so that their separation from each other takes a long time which results in the formation
of dark oblique solitons at a finite distance from the obstacle.

{\it 2. Convective and absolute instabilities. }
The presence of a positive imaginary part in the dispersion law $\omega = \omega(p)$
of the soliton oscillations
does not mean in itself that an arbitrary perturbation in a given point will
grow with time.
The stability of the soliton is determined by the asymptotic behavior of wave packets built from
harmonic waves. Let the Fourier transform of the initial density perturbation
be $\delta n_{p}$. Then the
time dependence of the perturbation is defined by the equation
\begin{equation}
 \delta n(t,y) \propto \int^{\infty}_{-\infty} \delta n_{p}e^{i[py-\omega(p)t]}dp \; .
 \label{perturbation}
 \end{equation}
Hence,  there are two possibilities.
In one case, the perturbation increases without limit at any fixed value of $y$.
This  situation corresponds to absolute instability.
In the other case the packet is carried away by the flow along the soliton plane so fast
that the perturbation tends to zero as $t \to \infty $ at any fixed value of $y$.
This  corresponds to convective instability.
We emphasize that from a practical point of view a convective instability
does not prevent the observation
of a finite size soliton in experiments and  does not violate
the stability of numerical simulations.

The asymptotic behavior of the perturbation (\ref{perturbation}) at $t \to \infty $
can be investigated by deformation of the contour of integration
in the complex $p$ plane.
The criterion distinguishing absolute and convective instabilities
can be formulated as follows (see \cite{ll-10}, Ch.VI): ({\it i})
in the complex $p$ plane the function
$p(\om)$  has values lying in upper and lower half-planes for
$\mathrm{Im}\,\om\equiv \om''\to+\infty$;
({\it ii}) when we decrease $\om''$ then for some values
of $\om$ the two values of $p$ moving from opposite
half-planes coincide in the branching point
$p_{br}$ of the function $p(\om)$;
({\it iii}) if these branching points correspond to values of $\om_{br}$
lying in the upper $\om$ complex half-plane, then we have absolute instability;
otherwise the instability is convective. The asymptotic behavior of the perturbation at
$t \to \infty $ is $\delta n(t,y) \propto e^{i(p_{br}y-\omega_{br}t)}/\sqrt{t}$,
where $p_{br}=p(\om_{br})$.

This investigation demands knowledge of the dispersion law $\omega(p)$ for all
values $p$ so that knowledge of just the long wave-length approximation
(\ref{gen}) is not sufficient for this aim.
Before going to concrete calculation, we have to make a following general remark.
The solution (\ref{1-2})
describes a soliton in the absence of a flow along its plane. In our problem the soliton
is created by the flow moving with respect the obstacle with velocity $M(=Mc)$.
Let  $\theta$ be the angle between a vector normal
to the soliton plane and the direction of the flow.
Then the normal component of velocity is $V=M\cos \theta $ and the flow has a
component $u =M\sin \theta$ along
the soliton plane.
(A solution of (\ref{1-1}) corresponding to this ``oblique'' soliton has been
obtained in \cite{egk-2006}.)
Since we want to investigate the stability of the soliton
in the ``laboratory'' coordinate frame with the obstacle at rest,
we have to transform the dispersion law
$\omega_0 (p)$ of the oscillation of a ``direct'' soliton (\ref{1-2})
to the ``laboratory'' reference frame which means that $\omega_0(p)$
should be replaced by $\omega (p)=u p+\omega_0 (p)$.

{ \it 3. Shallow dark solitons.}
The investigation of the nature of instability of the soliton (\ref{1-2})
is a quite complicated
problem. As a first step, we will study a soliton with a small amplitude,
which moves with the
near-sonic velocity $V$, $1-V \ll 1$. In this case the dispersion
law can be found analytically.
As is known, in the small amplitude limit the GP equation can
be reduced to the KP equation (see, e.g. \cite{kt-1988})
\begin{equation}\label{3-1}
    (\tn_t-\tn_x-\tfrac32\tn\tn_x+\tfrac18\tn_{xxx})_x=\tfrac12\tn_{yy}
\end{equation}
where $n= 1+\tn$ and we used the same notations $t,x,y$ for the new variables.

This equation has the soliton solution
\begin{equation}\label{3-2}
    \tn(x,t)=-s/\cosh^2[\sqrt{s}(x+Vt)]
\end{equation}
where $s=2(1-V)$.
Since in the small amplitude limit we have $1-V^2\cong2(1-V)$,
this solution is an approximation to the exact solution (\ref{1-2}).
Small oscillations of the soliton (\ref{3-2})
were studied in \cite{zakharov-1975,aps-1997}
where, for waves propagating with wave number $p$ along soliton,
the spectrum $\omega_0(p)=i\Gamma(p)$ was obtained,
which after transformation to the ``obstacle'' frame takes the form
\begin{equation}\label{3-7}
    \om=\om(p)=u p+i\Gamma(p)=u p+i({p}/{\sqrt{3}})\sqrt{s-{2p}/{\sqrt{3}}} \; .\\
\end{equation}

The above criterion can be easily applied to the dispersion relation (\ref{3-7}).
Indeed, at
$\om''\to+\infty$ this relation yields three values of $p$ one
of which ($p\propto -(\om'')^{3/2}$)
lies on the real axis, and the other two ($p\propto (\om'')^{3/2}e^{\pm\pi i/3}$)
are located in the upper and lower $p$ half-planes. Then,
in the branching point we have
$
    {d\om(p)}/{dp}=0
$
and this equation gives after a simple calculation the critical value of $p$:
\begin{equation}\label{3-8}
    p_{br}=[s+u^2-u\sqrt{u^2-s}]/\sqrt{3}.
\end{equation}
Substitution of this value into Eq.~(\ref{3-7}) yields the critical value of $\om$:
\begin{equation}\label{3-9}
    \om_{br}=p_{br}(\sqrt{u^2-s}+2u)/3.
\end{equation}
 For $u^2<s$ we get
$
    \om''_{br}=((s-u^2)/3)^{3/2}>0
$
which means the instability is absolute. However, $p_{br}$ and $\om_{br}$ have real values for
$
    u^2>s
$
and in this case the instability is convective. Thus, the transition
from absolute to convective instability takes place at
$u^2>s$, that is at
 $   M^2>1.$
The shallow oblique solitons are convectively unstable therefore for supersonic
flow of a BEC past an obstacle.

{\it 4. Convective instability criterion for deep dark solitons.}
In the case of deep dark solitons there is no analytical expression
for the full instability spectrum,
although its asymptotic expressions are known: for small $p\ll q$ we have (see \cite{kt-1988})
\begin{equation}\label{4-1}
    \Gamma^2(p)=k^2p^2/3-(2-k^2)p^3/(3\sqrt{3}),\quad p\ll q,
\end{equation}
where $k^2=1-V^2$ and
\begin{equation}\label{4-2}
    q=\left[-(1+V^2)+2\sqrt{V^4-V^2+1}\right]^{1/2}\;.
\end{equation}
Notice that in the limit $p \to 0$ Eq.~(\ref{4-1}) coincides with the general result
(\ref{gen}) obtained for the GP equation.
For $q-p\ll q$ we have (see \cite{psk-1995})
\begin{equation}\label{4-3}
    \Gamma^2(p)=q(q-p)/\beta(k),\quad q-p\ll q,
\end{equation}
where  $\beta(k)$ is evaluated numerically.
We approximate this behavior by an interpolating function
\begin{equation}\label{4-4}
    \Gamma^2(p)=f(p)(q-p)
\end{equation}
with $f(p)=a(k)p^2+b(k)p^3+c(k)p^4$,
where
\begin{equation}\label{4-7}
\begin{split}
    &a(k)=k^2/(3q),
    b(k)=[k^2/q-(2-k^2)/\sqrt{3}]/(3q),\\
    &c(k)=[3/\beta(k)-2k^2+q(2-k^2)/\sqrt{3}]/(3q^3),
    \end{split}\nonumber
\end{equation}
and $\beta(k)$ can also be approximated as
 $   \beta(k)=3/[(1+\sigma k^2)k^2],\quad \sigma\cong 0.596$.
This interpolating expression yields the dependence of $\Gamma$ on $p$ shown in Fig.~2 for
several values of $V$.
\begin{figure}[bt]
\includegraphics[width=6.5cm,height=4cm,clip]{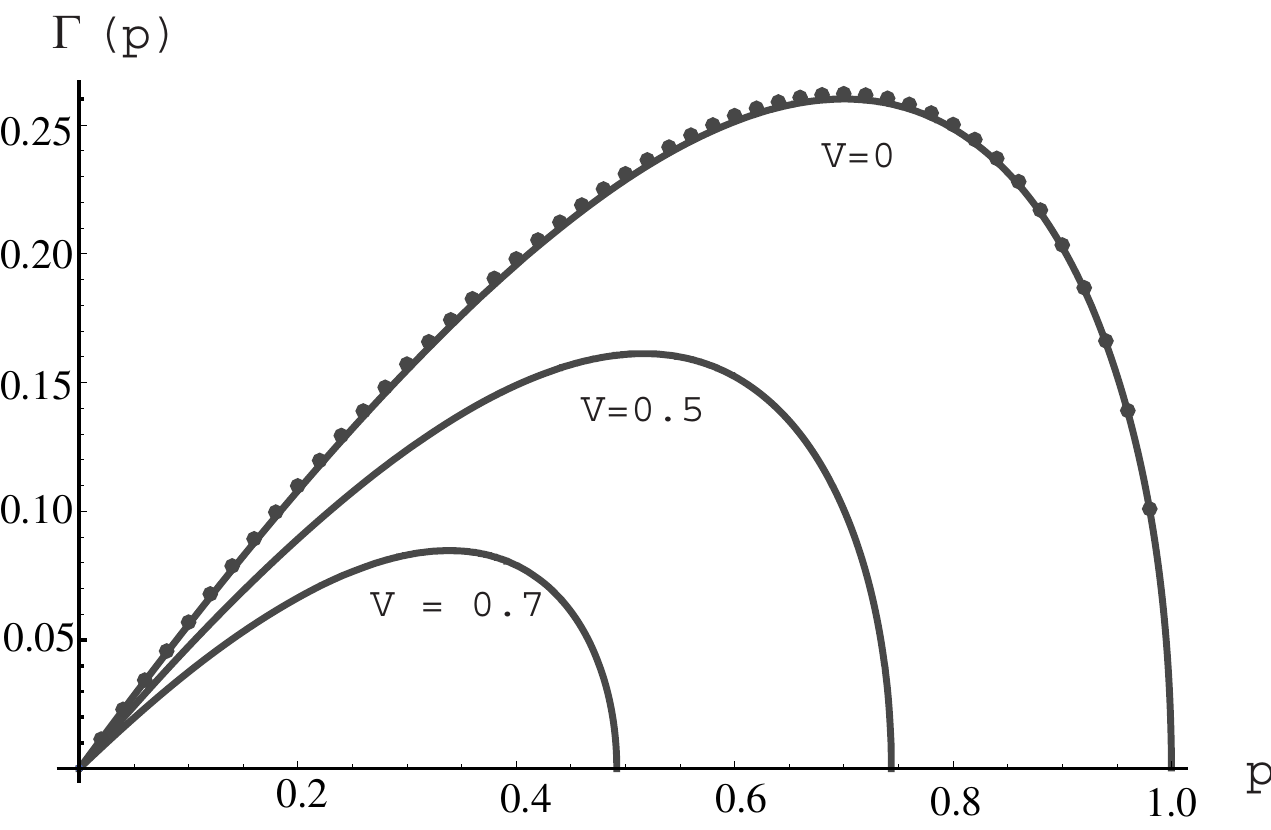}
\caption{Growth rate approximated by Eq.~(\ref{4-4}) with $f(p)$ given by
the interpolation formula as a function of wave vector $p$ of a harmonic wave
propagating along the oblique dark solitons with $V=0.7,\,V=0.5,\,V=0$.
The exact dependence found numerically for $V=0$ is shown by dots for illustration.}
\label{fig2}
\end{figure}

On the contrary to the shallow soliton case, an expression for the critical
flow velocity $M_{cr}$
cannot be given in explicit form. Therefore we shall follow Sturrock's
\cite{sturrock-1958} formulation of the
criterion and notice that the function $p=p(\om)$ determined implicitly by the equation
 $   \om(p)=u p +i\Gamma(p,V)$,
changes its behavior at $u=u_{cr}$: for $u<u_{cr}$ it is represented
in the complex $p$ plane by disconnected curves, whereas for $u>u_{cr}$
these curves are connected
with each other. Therefore a ``space-like'' Fourier representation
of a wave packet of disturbance
cannot be transformed to its ``time-like'' representation for $u<u_{cr}$ and in this case
the instability is absolute. For $u>u_{cr}$ we can deform the
contour $p$ so that a single-valued function
$\om(p)$ is defined for all $p>0$ and hence a ``space-like'' packet
is also a ``time-like'' packet
which means that the instability is convective (see \cite{sturrock-1958} for details).
At the branching point of $p=p(\om)$ we have $d\om/dp=0$, that is the values
$p_{br}$ satisfy the equation
\begin{equation}\label{4-10}
    u=-id\Gamma/dp \equiv -i\Gamma'(p,V),
\end{equation}
where $\Gamma(p,V)$ has either real or purely imaginary values for real $p$.
Hence, the critical value of $u$ corresponds to the appearance of a double
zero $p_{br}$ of the equation
(\ref{4-10}) on the real axis of $p$ as it takes place in the above considered case
of shallow solitons, that is $dp_{br}/du=\infty$ at $u=u_{cr}$. (In other words,
$p_{br}$ has here a branching point as a function of $u$.) Differentiation of Eq.~(\ref{4-10})
with respect to $u$ then gives  the equation
 $\left.{d^2\Gamma}/{dp^2}\right|_{p=p_{cr}}=0$
for the corresponding critical value $p_{cr}(V)$. Substitution of $p_{cr}$, found in this way,
into squared equation (\ref{4-10}) yields with account of  (\ref{3-7}) the equation
\begin{equation}\label{4-12}
 M_{cr}^2\sin^{2}\theta=-\left[\Gamma'\left(p_{cr}(M_{cr}\cos \theta),M_{cr}\cos \theta \right)
  \right]^2
\end{equation}
which determines implicitly the critical value $M_{cr}$ of the
transition from absolute instability to
a convective one as a function of the angle $\theta$.
For $\Gamma(p)$ given by Eq.~(\ref{3-7}) these
formulae reproduce the above results derived for shallow solitons. The plot of
$M_{cr}(\tan \theta)$
calculated for the instability spectrum defined by the interpolation
formula (\ref{4-4}) is shown in Fig.~3 by a solid line.
As we see, the curve obtained with the use of the interpolation formula agrees well enough with
the results derived from the numerically calculated instability spectrum.
\begin{figure}[bt]
\includegraphics[width=6.5cm,height=4cm,clip]{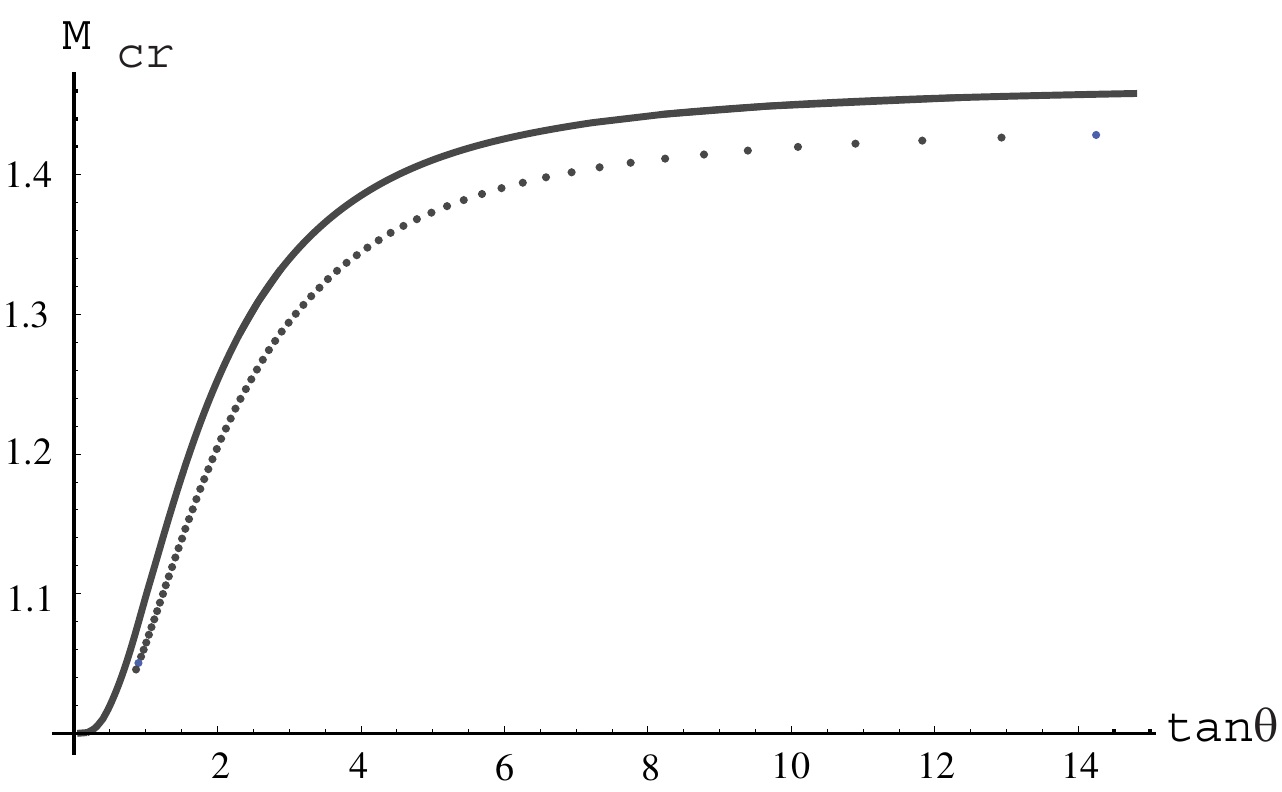}
\caption{The boundary between regions of convective and absolute instabilities
in the plane of parameters ($\tan \theta, M_{cr}$). Below this curve the instability is
absolute and above it it is convective. Dots correspond to the transition points
calculated from a numerical solution of the eigenvalue problem
for the instability spectrum.}
\label{fig3}
\end{figure}

In experiment \cite{cornell-2005}, the obstacle was created by a laser beam. Hence
the flow of BEC past a cylindrical obstacle was actually studied. Our results can be
applied to this case without any changes due to the isotropy of the growth rate spectrum
$\Gamma(p)$, $p=(p_x^2+p_y^2)^{1/2}$, in momentum space. Due to this isotropy,
the unstable modes grow up with the same increments in all directions and if the addition
of convection can remove a wave packet built from unstable modes from a region
finite in the  $x$ direction, then it removes it from a region finite in the  $y$ direction, too.

{\it 5. Conclusion.} Thus, for $M<1$ oblique solitons are absolutely unstable with respect to
the transverse instability leading to their decay to vortex pairs of
opposite polarities. However, for supersonic flow $M>1$
a region of angles $\theta$ appears where oblique solitons are only convectively unstable
which provides the possibility of
formation of oblique solitons generated by the flow of a BEC past an obstacle.
For $M$ greater than some maximal value
$M_{max}\cong 1.46$, oblique
solitons are only convectively unstable  for any angle $\theta$ inside the Mach cone.
These results explain the stability of oblique solitons observed in numerical
simulations \cite{egk-2006}. We suppose that this method of generation of dark solitons
in a BEC presents new possibilities for their experimental study.

We thank A. Gammal and Yu. G. Gladush  for help with numerical calculations. Discussions with
G. Bruun, G. A. El, V. A. Mironov, P. Pedri and L. A. Smirnov are appreciated.
Work of A.M.K. was supported by RFBR.

\end{document}